\documentclass{mn2e}

\usepackage{amssymb}
\usepackage{astrobib}
\usepackage{epsfig}
\usepackage{latexsym}
\usepackage{times}

\def\i{\item}

\newcommand{\bei}{\begin{itemize}}
\newcommand{\eei}{\end{itemize}}
\newcommand{\bef}{\begin{figure}}
\newcommand{\eef}{\end{figure}}
\newcommand{\ben}{\begin{enumerate}}
\newcommand{\een}{\end{enumerate}}
\newcommand{\beq}{\begin{equation}}
\newcommand{\eeq}{\end{equation}}
\newcommand{\ber}{\begin{eqnarray}}
\newcommand{\eer}{\end{eqnarray}}

\newcommand{\bb}{\bf B}
\newcommand{\nb}{\bf \nabla}

\newcommand{\gcc}{\mbox{${\rm g} \, {\rm cm}^{-3}$}}
\newcommand{\mdot}{\mbox{$\dot{M}$}}
\newcommand{\msun}{\mbox{{\rm M}$_{\odot}$}}
\newcommand{\pdot}{\mbox{$\dot P$}}

\newcommand{\dmdt}{{\mbox{{\rm M}$_{\odot}$}} {\rm yr}$^{-1}$}

\newcommand{\lsim}{\raisebox{-0.3ex}{\mbox{$\stackrel{<}{_\sim} \,$}}}
\newcommand{\gsim}{\raisebox{-0.3ex}{\mbox{$\stackrel{>}{_\sim} \,$}}}

\title[{Magntic Fields of MSPs}]
{The Magnetic Fields of Millisecond Pulsars in Globular Clusters}
\author[Konar]
{ Sushan Konar\\ 
Physics, Harish-Chandra Research Institute, Allahabad 211019, India \\ 
e-mail : sushan@hri.res.in}

\begin{document}

\date{}

\pagerange{\pageref{firstpage}--\pageref{lastpage}} \pubyear{2009}

\maketitle

\label{firstpage}

\begin{abstract}
Many of the characteristic properties of the millisecond pulsars found
in globular clusters are markedly different from those in the Galactic
disc. We find that one such physical parameter is the surface magnetic
field strength.   Even though the  average spin-periods do  not differ
much the  average surface  magnetic field is  2-5 times larger  in the
globular cluster pulsars.  This  effect could be apparent, arising due
to  one   or  more  of  several  biases.    Alternatively,  if  future
observations  confirm this  effect  to  be real,  then  this could  be
interpreted as  a preferential recycling of pulsars  in tight binaries
where the mass transfer takes place at high accretion rates.
\end{abstract}


\section{introduction}
\label{sec01}

A  radio  pulsar  is  a  strongly magnetized  rotating  neutron  star.
Discovered    serendipitously~\cite{hews68},     the    pulsars    are
characterized  by  their  short  spin-periods  ($P$)  and  very  large
inferred surface magnetic fields ($B$).  However, the detection of the
1.  5~ms pulsar  B1937+21~\cite{back82} heralded a new  genre, that of
the radio  millisecond pulsars (MSP).  The ranges  of the spin-periods
and the surface  magnetic fields of these MSPs place  them in a nearly
disjoint  region of  the $B-P$  plane from  the normal  radio pulsars.
Close to two thousand radio pulsars have been detected to date with $P
\sim 10^{-3}  - 10$~s and $B  \sim 10^8 -  10^{15}$~G.  Amongst these,
the MSPs are  typically characterized by $P \lsim  30$~ms, and typical
surface field  strengths of $\sim  10^8 - 10^9$~G  with characteristic
spin-down ages of $\sim 10^9$~yr.

The  connection between  these two  seemingly disjoint  populations of
pulsars  is realized  through the  binary association  of  the neutron
stars. An MSP  is understood to descend from  an ordinary, long-period
pulsar that  have been spun-up  and recycled back  as an MSP in  a low
mass  X-ray  binary (LMXB)  by  mass  accretion~\cite{rad82}.  In  the
particular case  of an isolated MSP,  like PSR B1937+21,  the donor is
later  destroyed probably  by the  wind of  the newly  recycled pulsar
itself~\cite{alpr82}.  In  general the MSPs  are considered to  be the
end products  of LMXBs and intermediate-mass X-ray  binaries where the
primary neutron star is assumed to have formed by the core-collapse of
a massive ($M \gsim 8 \msun$) star~\cite{bisn74,bhat91}.

This  scenario   for  MSP  formation   has  been  backed   by  several
observational indications  over the years.  The  strongest support for
the connection between MSPs and X-ray binaries come from the discovery
of     coherent     millisecond     X-ray    pulsations     in     SAX
J1808.4~\cite{chak98,wijn98}   and    subsequently   in   many   other
systems~\cite{wijn05}.   These  accreting  millisecond  X-ray  pulsars
(AMXP) are understood to be the immediate precursors of the radio MSPs
and are expected to turn  into those when active mass accretion stops.
The identification of PSR J1023+0038 as an MSP which has only recently
lost its  accretion disc~\cite{arch09} provides almost  a direct proof
of this theory of MSP formation.

However,  it  seems that  the  large  number  of pulsars  detected  in
globular  clusters  in  the   recent  years  have  somewhat  different
characteristics than  the pulsars observed  in the Galactic  disc.  To
begin  with, the  fraction of  binary and  millisecond pulsars  in the
clusters are $\sim 40\%$ and  $\sim 92\%$ respectively, whereas in the
disc these fractions are only $\sim  5\%$ and $\sim 4\%$.  The size of
the MSP  population is much larger  in the clusters than  in the disc.
In fact, the globular clusters (about  150 of them) known to orbit the
Milky Way contain roughly three orders of magnitude more observed MSPs
per  unit  mass than  the  Galactic  plane,  which contains  73  known
MSPs~\cite{caml05}.  The  conditions prevailing in  a globular cluster
is rather  different from that in  the Galactic disc  primarily due to
the  extremely high  stellar densities  in the  clusters.  One  of the
obvious  effect of this  high density  is a  dramatic increase  in the
number of binaries, as well as in the rate of close stellar encounters
allowing   for   many  different   channels   for  binary   formation.
Effectively, these systems provide different ways of pulsar recycling,
than seen in  the disc where the MSPs are  mainly formed in primordial
low-mass binaries.

Therefore,  we  can  logically expect  the  MSPs  in  the disc  to  be
different  from the MSPs  in the  clusters. This  is indeed  the case.
There is a  greater proportion of single MSPs in  the clusters and the
majority of the  binary MSPs have very short  orbital periods compared
to those in  the Galactic disc. In fact, many  of the cluster binaries
have properties  similar to  those of the  rare eclipsing  black widow
pulsars  seen in  the  Galactic disk  population~\cite{king03,frei05}.
Recently,  Bagchi \& Ray~\citeyear{bagc09a,bagc09b}  have investigated
the effect  of different  types of stellar  collisions on  the orbital
parameters of the binary MSPs  in globular clusters. They too find the
cluster MSPs to be significantly different from those in the disc.

In this work, we look at  one of the important intrinsic parameters of
the MSPs, namely the surface magnetic  field. We find that a subset of
slower MSPs in the globular clusters (for which field measurements are
available) appear to  have surface magnetic fields that  are 2-5 times
larger  compared   to  their  disc  counterparts.    Even  though  the
spin-periods  of this  subset of  cluster pulsars  are similar  to the
spin-periods  of the MSPs  in the  disc. In  order to  understand this
fact, we look at the  spin-period and the magnetic field distributions
of the MSPs in the disc  and in the globular clusters and compare them
to  check for implied  similarities/differences in  their evolutionary
histories.  We also look at  the details of the field evolution itself
to see  if the physics  of the  MSP formation in  the disc and  in the
clusters  could actually give  rise to  different surface  fields.  We
examine   whether  close  stellar   encounters  resulting   in  binary
disruption and/or  re-formation of  new binaries could  be responsible
for the higher magnetic fields observed in cluster MSPs. Or if this is
an apparent effect caused by other external factors.

To this end, first we discuss the MSP statistics in Sec.2. In Sec.3 we
look  at the  details of  the field  evolution and  the effect  of the
cluster  conditions  on  the  evolution.  Finally,  we  summarize  our
conclusions in Sec.4.

\setcounter{figure}{0}
\begin{figure*}
\epsfig{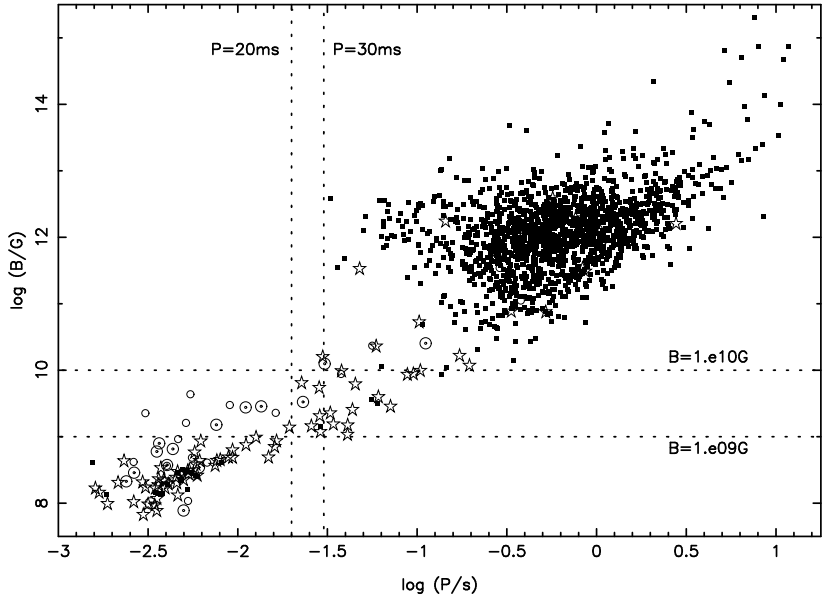}
\caption[]{All   known   Galactic  pulsars,   with   a  reliable   $B$
measurement, in  the $B-P$ plane.  The isolated and binary  pulsars in
the  Galactic  disc  are  marked  by filled  squares  and  open  stars
respectively.   The  isolated  and  binary  pulsars  in  the  globular
clusters are marked by open  circles and circles with dots within. The
data is from the ATNF on-line catalog, taken in April 2010.}
\label{fig01} 
\end{figure*}

\section{the millisecond pulsars}
\label{sec02}

\subsection{the definition}

It is  well known  that the MSPs  occupy a distinctly  separate region
from the normal pulsars in the $B-P$ plane.  However, defining the MSP
population  with  some accuracy  is  not  simple  since this  requires
separating  out  a  class  of  neutron stars  that  have  undergone  a
particular kind of binary  evolution.  The primary observed quantities
of  a pulsar  are  $P$ and  $\pdot$,  and the  most important  derived
quantity using  both $P$ and $\pdot$  is the dipolar  component of the
surface magnetic field given by~\cite{manc},
\beq
B_s \simeq 3.2 \times 10^{19} (P\pdot)^{1/2}~{\rm G},
\label{eq01}
\eeq
where  $P$ and  $\pdot$ are  in units  of second  and  $ss^{-1}$. Even
though  the  term  {\em  millisecond pulsar}  has  traditionally  been
reserved  for  recycled pulsars  with  ultra-fast  rotation ($P  \lsim
30$~ms)  and a  low magnetic  field ($B  \simeq 10^8  -  10^9$~G), the
definition has  mostly used  the condition on  $P$.  Using  $\pdot$ to
classify pulsars is somewhat problematic because for a large number of
pulsars  (particularly for  those in  the clusters)  a  measurement of
$\pdot$ is  either not available or  the value is not  reliable due to
contamination from the proper motion of the pulsar.

The most reasonable  criterion to define an MSP should  be to use both
$P$  and $\pdot$  since  the  evolution of  these  two parameters  are
inter-related  in the  formation  process of  MSPs.   Some effort  has
indeed  been made  to use  such  a relation  assuming the  MSPs to  be
recycled through straightforward LMXB evolution~\cite{stor07}. But the
situation  is different  in globular  clusters  where, due  to a  much
larger probability of stellar collisions, it is possible for an MSP to
have  gone  through  very  complex binary  evolution.   Therefore,  in
general, it is  not easy to separate the MSPs  from the normal pulsars
using a definitive relation involving both $P$ and $\pdot$.

\bef 
\epsfig{file=fig02.ps,width=165pt,angle=-90}
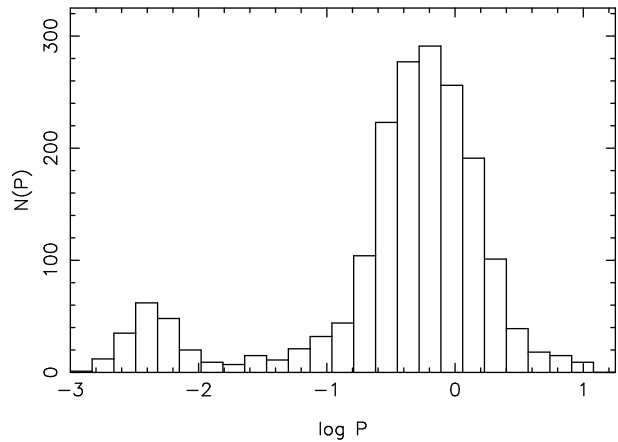
\caption[]{Histogram showing the distribution of spin-periods ($P$) of
all known  pulsars.  The data  correspond to the ATNF  on-line catalog
and Paulo  Freire's on-line catalog of globular  cluster pulsars, April
2010.}
\label{fig02} 
\eef

In Fig.\ref{fig01} we have plotted  all known pulsars (with a reliable
estimate for $B$) in the $B-P$  plane, marking some of the $B$ and $P$
lines relevant for separating  out the MSP population.  Unfortunately,
these simple  criteria run  into trouble with  objects like  the 16~ms
pulsar J0537-6910  in LMC~\cite{mars98}.  As this pulsar  has a strong
strong  magnetic field  ($B \sim  10^{11}$) and  is also  rather young
($\tau \sim 10^4$yrs), it is not  likely to be a member of the generic
recycled MSP population (this extra-Galactic pulsar is not included in
Fig.\ref{fig01}).  Therefore as a  working hypothesis we shall use the
classical definition  of $P  \ \lsim  \ 30$~ms (  $\log (P/s)  \ \lsim
-1.5$ ),  barring such obvious  misfits like J0537-6910.  This  way of
defining MSPs does appear to have a certain merit.  In Fig.\ref{fig02}
a  histogram of  the spin-period  of all  known pulsars  is  made.  It
appears that the pulsars with $P \ \lsim \ 30$~ms may indeed represent
a separate class as the period histogram shows a sharp dip around this
value of $P$.

\subsection{the statistics}

\begin{table}
\begin{tabular}{|l|r|r|r|} \hline
MSP Population & $P_{\rm av}$ &  $B_{\rm av}$ & $P^{\rm B}_{\rm av}$ \\ 
($P \leq 30$ms) & ms & $10^8$~G & ms \\ 
&&& \\ 
Galactic Disc :   && \\  
isolated & 5.82 (19) & 3.11 (18) & 5.97 (18) \\
binary   & 8.03 (54) & 9.32 (52) & 8.45 (52) \\
all      & 7.75 (73) & 7.72 (70) & 7.82 (70) \\
&&& \\ 
Globular  Clusters :  &&& \\
isolated & 5.82 (61)  & 14.70 (11) & 6.16 (11) \\
binary   & 5.58 (68)  & 18.84 (14) & 8.62 (14) \\
all      & 5.70 (129) & 17.02 (25) & 7.54 (25) \\
&&& \\ \cline{1-4}
\end{tabular}
\caption{The MSP Statistics.  In the  first column the type of the MSP
sub-population has  been indicated.  The  second and the  third column
show  the average  values  of the  measured  spin-periods and  derived
dipolar  field strengths  respectively.  $P^{\rm  B}_{\rm av}$  in the
fourth  column  is the  average  period  calculated  using only  those
pulsars for which $B$ measurements  exist.  The spin-periods are in ms
and  the  magnetic  fields  are  in  $10^8$~G.   The  numbers  in  the
parentheses denote  the numbers  of objects available  for calculating
the averages.   The data is from  the ATNF on-line  pulsar catalog and
Paulo Freire's catalog of globular cluster pulsars, April,2010.}
\label{tab01}
\end{table}

Interesting facts emerge when the MSP population, as defined above, is
subjected to  statistical analysis.  To  begin with, we  separate them
into  two groups -  MSPs residing  in the  globular clusters  (GC) and
those in  the Galactic  disc (GD).  Each  of these groups  are further
divided into isolated and binary pulsars.  Table.\ref{tab01} shows the
average values  of $P$ and $B$  for each of  these sub-populations. It
needs to be noted that the  number of pulsars with a measured value of
$B$ can be  much smaller than the actual number of  pulsars in a given
sub-group (see the  corresponding numbers in Table.\ref{tab01}).  This
is particularly true of the  cluster pulsars.  Because of this we have
calculated  another average  of $P$  using only  those pulsars  with a
known $B$.  This average, $P^{\rm B}_{\rm av}$, is expected to be more
commensurate with  $B_{\rm av}$ than  the $P_{\rm av}$  obtained using
all  the pulsars  in the  sub-population.  The  significance  of these
quantities shown in Table-\ref{tab01} are discussed below.

\setcounter{figure}{2}
\begin{figure*}
\epsfig{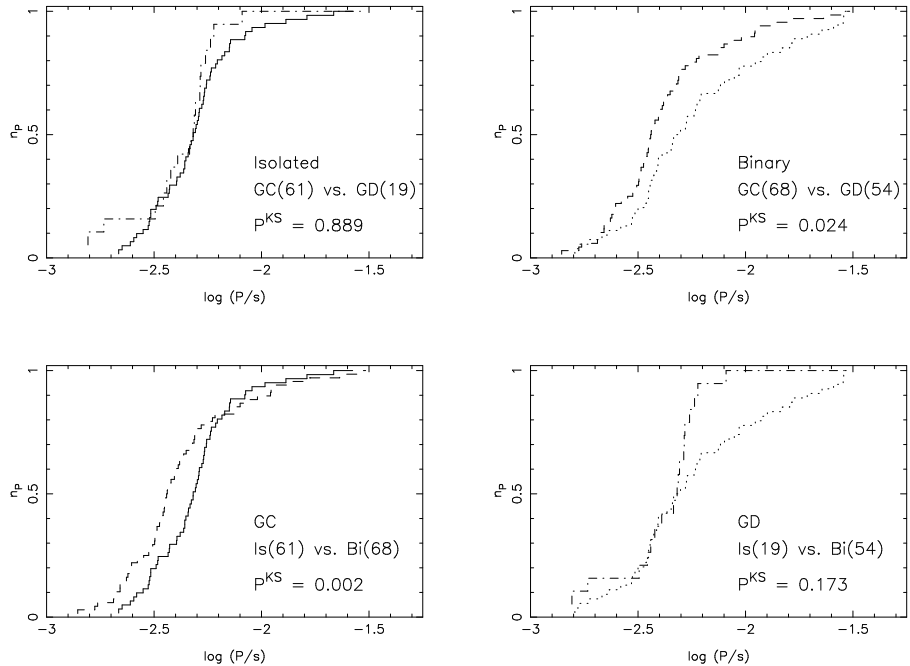}
\caption[]{Kolmogorov-Smirnov   test  for  the   $P$-distributions  of
different   subgroups  of   the  MSPs.    The   cumulative  fractional
distributions are shown for easy comparison. The solid, and the dashed
lines correspond to  the isolated and the binary  MSPs in the globular
clusters; whereas  the dash-dotted and the dotted  lines correspond to
the isolated and the binary MSPs  in the Galactic disc.  The number of
objects in  each subgroup are  shown within the parentheses.   The K-S
probabilities are indicated for each test.}
\label{fig03} 
\end{figure*}

\subsubsection{Spin period}

 The well-known fact that  the MSPs in globular
 clusters   are,  in   general,  spinning   faster  than   their  disc
 counterparts    is   immediately    seen   from    the    values   in
 Table-\ref{tab01}.   However, it can  also be  seen that  the $P_{\rm
 av}$ for almost  all the subgroups are similar  except for the binary
 MSPs in the disc which has a larger average spin-period than the rest
 of the subgroups.  In other  words, the disc binary MSPs are spinning
 more  slowly  than  the  rest  of  the  MSPs.   However,  an  average
 difference of $\sim$3~ms may not indicate anything significant as the
 standard deviation of  $P$ in these sub-groups range  from $\sim$ 4 -
 7~ms.  Moreover, it is not clear if  this could be due to any kind of
 observational bias.  Though  it should be noted that  the size of the
 sub-groups are more  or less similar except for  the isolated objects
 in the disc (about a factor  of 3 smaller).  Therefore a bias (of any
 kind) would be stronger for the isolated disc MSPs more than the rest
 of the groups.

In order to understand the relative nature of the $P$-distributions of
the  various  sub-populations, beyond  the  simple averages  discussed
above  we  have  performed  the  Kolmogorov-Smirnov  tests  on  these
populations. In Fig.\ref{fig03} we compare the cumulative fractional
$P$-distribution corresponding to each K-S test. And our conclusions
from these tests are summarized below.
\ben
\i The isolated and the binary MSPs in the globular cluster itself are
the least  correlated ($P_{KS} \sim  0.002$) and is in  good agreement
with  that found by  \citeN{hess09}. The  result itself  is surprising
because these  two subgroups  have very similar  average spin-periods.
Also the  luminosity distributions of  isolated and binary  GC pulsars
have been  found to  be statistically similar~\cite{hess07}.   The low
correlation is  also contrary to expectations because  all the pulsars
in  the clusters are  likely to  have similar  evolutionary histories,
strongly influenced by stellar  encounters. Since it is quite possible
for an isolated MSP to acquire a companion or a binary MSP to lose one
given the high rates of stellar collisions in a globular cluster which
effectively means  that in a  globular cluster the phase  (isolated or
binary) in  which an MSP  is observed at  a given time could  be quite
temporary.
\i The isolated and the binary  MSPs in the galactic disc are also not
correlated ($P_{KS}  \sim 0.2$) indicating major  differences in their
evolutionary  histories.  This is  supported, to  some extent,  by the
fact that  their average  spin periods are  different (but not  by any
significant amount, as mentioned earlier).
\i Similarly, there is little correlation ($P_{KS} \sim 0.03$) between
the spin-distributions  of the binary  populations in the disc  and in
the  clusters.  The  explanation for  this absence  of  correlation is
possibly due to the fact that  in the disc the MSP recycling typically
happens in  primordial binaries whereas  in the clusters  the binaries
could  well have  formed  by recent  stellar  encounters.  Hence,  the
nature of the associated MSPs could be very different.
\i Another  surprising result is the very  strong correlation ($P_{KS}
\sim 0.9$) between the the  spin-distributions of isolated MSPs in the
disc and in  the clusters.  The average spin-periods  are similar too,
which may indicate similar evolution  of the isolated MSPs in the disc
and the clusters.  But these two populations are not expected to be so
well correlated.   In the disc the  the isolated MSPs  are supposed to
evolve in primordial LMXBs and then evaporate their companions. In the
clusters, this is certainly  a possibility. However, the isolated MSPs
are more  likely to  be results of  stellar encounters  and consequent
disruption of binaries in the  clusters.  It should also be remembered
that the size of one of  the subgroups (the disc pulsars) in this test
is much smaller  than the other and this could  introduce some bias in
the K-S test.
\een

\setcounter{figure}{3}
\begin{figure*}
\epsfig{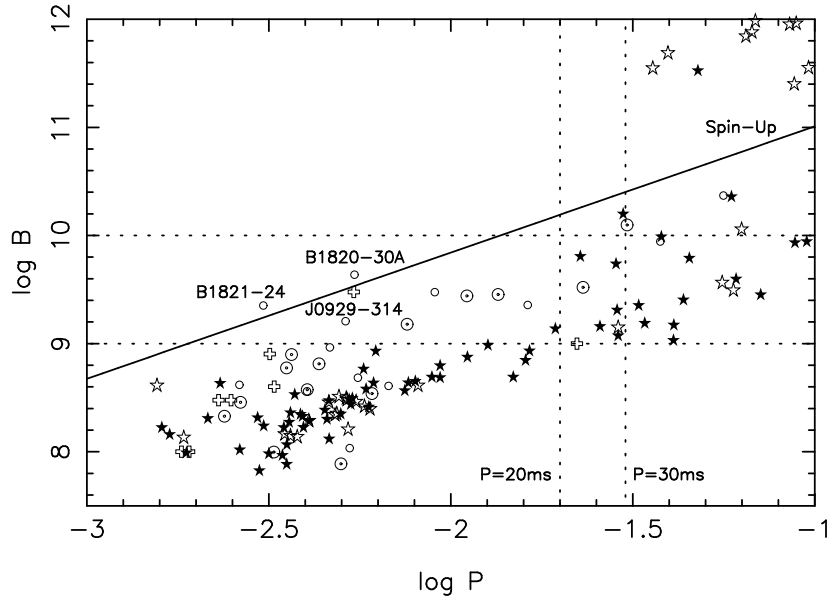}
\caption[]{All radio  MSP candidates (left  of the $P=30$~ms  line) in
the $B-P$ plane, along with the AMXPs (only estimated upper limits for
$B$).  The isolated and binary pulsars in the Galactic disc are marked
by  open and  filled  stars respectively.   Whereas  the isolated  and
binary pulsars in the globular clusters are marked by open circles and
circles with dots  within. The AMXPs are marked  by open crosses.  The
line marked  {\bf Spin-Up} corresponds to  Eq.\ref{eq04}, defining the
minimum spin-frequency attainable by recycling.  The radio MSP data is
the same as  that in Fig.\ref{fig01}. The AMXP data  is taken from the
references cited in Table-\ref{tab02}. }
\label{fig04} 
\end{figure*}

\subsubsection{Surface magnetic field} 

First of all, it  should be noted
that  the $B$  measurement is  available only  for a  small  number of
pulsars in the globular clusters (25  out of a total of 129).  This is
because of the difficulty in measuring the $\pdot$ there due to proper
motion contamination.   In fact, in many  cases $\pdot$ is  seen to be
negative even when  the system is non-accreting (and  therefore has no
reason to  spin-up) indicating that the  measured value is  not at all
reliable.  In such  cases there is no way  to estimate $B$. Evidently,
it would be  relatively easier to measure larger  $\pdot$s.  This then
automatically introduces a bias,  that of preferentially measuring $B$
in systems  that have higher field  values.  On the  other hand almost
all the  MSPs in  the disc  have a reliable  $B$ measurement.   So any
comparison  between the  disc  and  the cluster  MSPs  suffer from  an
inherent bias in this respect.

Given  the above  problem  it is  still  interesting to  see that  the
magnetic fields of the MSPs in  clusters are larger by a factor of 2-5
compared to their  disc counterparts.  Unlike in the  case of $P$ this
difference is much larger than the standard deviation in $B$.  Yet the
corresponding $P^{\rm B}_{\rm av}$s are similar in the disc and in the
clusters, both for the isolated and the binary MSPs. However it should
be noted that  the fraction of MSPs, with  field measurements, is much
smaller in  the clusters compared to the  disc (see Table\ref{tab01}).
For example, the  fractions of MSPs with field  measurements are 11/61
(isolated)  and 14/68  (binary)  in the  clusters.   In contrast,  the
corresponding numbers  in the disc  are 18/19 and  52/54 respectively.
Consequently, in  the disc $P_{\rm  av}$ and $P^{\rm B}_{\rm  av}$ are
very  similar for  both the  isolated and  the binary  pulsars, which,
however, is not the case in the clusters.  It can be seen that $P^{\rm
B}_{\rm av}$  is somewhat  larger than $P_{\rm  av}$ for  the isolated
cluster pulsars whereas  $P^{\rm B}_{\rm av}$ is much  larger than the
corresponding   $P_{\rm  av}$   for  the   binaries   indicating  that
$B$-measurements  are  probably  being  selectively  made  for  slower
pulsars in the  clusters.  Keeping this in mind,  we can then conclude
that the  subset of slow  cluster pulsars, for  which $B$-measurements
are available, appear  to have higher magnetic fields  compared to the
disc pulsars  that have  on the average  similar spin-periods  to this
subset.  Therefore  future measurements of the magnetic  fields of the
faster pulsars in the clusters  are required to make proper comparison
of these two sets.

In Fig.\ref{fig04} we  have plotted all the radio  MSPs (as defined in
Sec.2)  in  the  B-P  plane,  along  with  the  AMXPs  for  which  $B$
measurements  (mostly   order  of  magnitude  of   upper  limits)  are
available.  It is seen that mostly,  for a given value of $P$, the MSP
with the highest value of $B$ is a cluster pulsar.  During the process
of recycling a pulsar is spun  up in the accretion phase.  The maximum
spin-up, for  a given rate  of accretion and  a given strength  of the
surface   field,    is   given   by   the    following   relation   of
spin-equilibrium~\cite{alpr82,chen93}:
\ber
P_{\rm eq}  
&\simeq&  1.9 \,  {\rm ms}  
          \left(\frac{B}{10^9~{\rm G}}\right)^{6/7} 
          \left(\frac{M}{1.4  \msun}\right)^{-5/7} \nonumber \\
&&  \; \;  \;  \;  \; \;  \;  \, \times  
          \left(\frac{\mdot}{\mdot_{\rm Ed}}\right)^{-3/7} 
          \left(\frac{R}{10^6~{\rm cm}}\right)^{16/7} \ .
\label{eq04}
\eer
Here $M$  \& $R$ denote the mass  and the radius of  the neutron star,
$\mdot$ \&  $\mdot_{\rm Ed}$  stand for the  actual and  the Eddington
rate of  mass accretion and $B$  is the surface  magnetic field.  When
$\mdot$ equals  $\mdot_{\rm Ed}$ the  above relation defines  the {\em
spin-up} line i.e,  the minimum period to which a  pulsar with a given
magnetic field can be spun-up.   However, even though all disc pulsars
are consistent with the spin-up condition, some of the cluster pulsars
like B1821-24 or B1820-30A appear to  be above the spin-up line in the
$B-P$  plane.    Given  the  errors  and  uncertainties   in  the  $B$
measurement, even if  these pulsars do not actually  violate the above
mentioned condition  it can  be said with  some confidence  that their
field  values are rather  high, particularly  in comparison  with the
disc MSPs with similar $P$.  

However,  there is  a  serious caveat  to  this. Even  among the  disc
pulsars, only a few (8 out of 18 isolated and 30 out of 52 binary MSPs
with  a $B$  measurement) have  $\pdot$  and hence  $B$ corrected  for
proper motions.   And in those  cases where such corrected  values are
available it  is seen that the un-corrected  values are systematically
larger. For example, the average of the corrected $B$ values for the 8
isolated  disc MSPs  is $2.4  \times 10^8$~G  whereas  the uncorrected
average is $2.8 \times 10^8$~G.  The corresponding numbers in the case
of binary  disc MSPs are $4.7  \times 10^8$~G and  $5.4 \times 10^8$~G
respectively.   Even  though the  errors,  introduced  in the  $\pdot$
measurement  by the proper  motion contamination,  are expected  to be
random it appears that in the  case of the disc pulsars the $B$ values
uncorrected  for the proper  motion tend  to be  systematically larger
than  the corrected values.   Admittedly, the  difference is  not very
large for  the disc pulsars.  But  in case of the  clusters, where the
proper  motion is  quite large,  this  difference could  also be  much
larger.  Therefore,  this may  make the measured  field values  in the
clusters systematically  larger.  Still, it is not  clear whether this
effect  is  enough to  explain  the  observed  difference between  the
average field values of the disc and cluster pulsars.

\setcounter{figure}{4}
\begin{figure*}
\epsfig{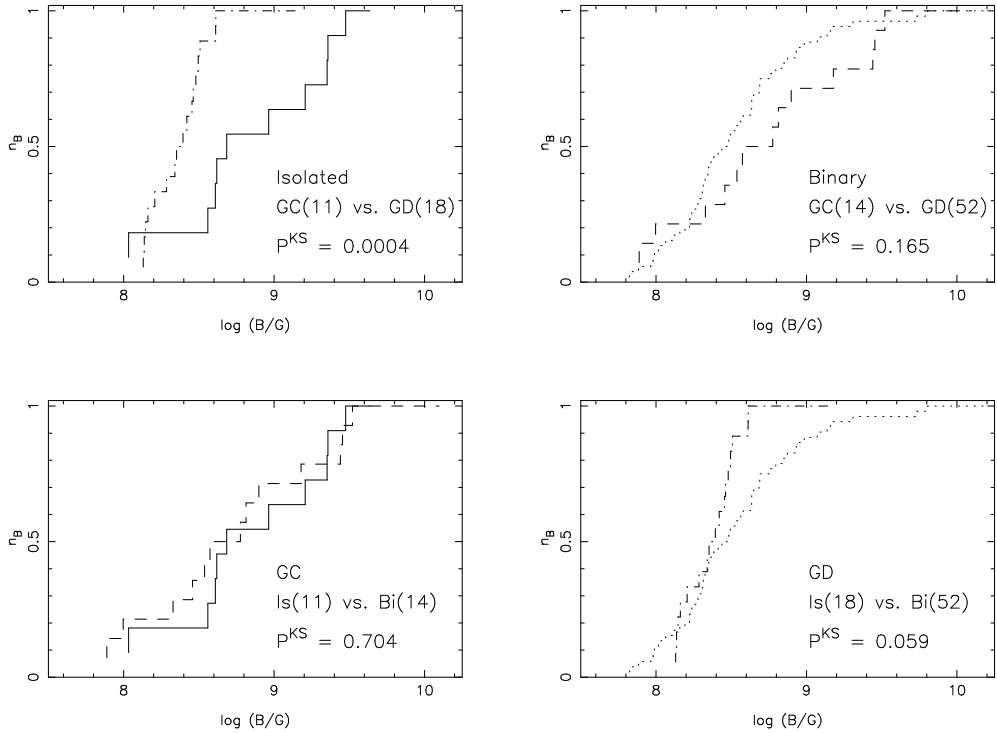}
\caption[]{Kolmogorov-Smirnov   test  for  the   $B$-distributions  of
different subgroups of the  MSPs, along with the cumulative fractional
distributions.  The solid,  and  the dashed  lines  correspond to  the
isolated and  the binary  MSPs in the  globular clusters;  whereas the
dash-dotted and  the dotted lines  correspond to the isolated  and the
binary  MSPs in  the Galactic  disc.  The  number of  objects  in each
subgroup are shown within  the parentheses.  The K-S probabilities are
indicated for each test.}
\label{fig05} 
\end{figure*}

Once again,  to understand the nature of  the $B$-distributions better
we  have performed  the K-S  tests  amongst the  various subgroups  of
pulsars. However, except for the binary  MSPs in the disc the sizes of
the other subgroups are really  small and hence the reliability of K-S
test results  are somewhat doubtful.  We summarize our  findings below
and the corresponding plots have been shown in Fig.\ref{fig05}.
\ben
\i The binary MSPs in the disc and the clusters have a low correlation
($P_{KS} \sim  0.2$), as do  the isolated and  the binary MSPs  in the
disc ($P_{KS} \lsim 0.1$). These  results are in conformity of what we
have seen  for the respective  $P$-distributions. And we  can conclude
that  these  subgroups   have  different  evolutionary  histories,  as
discussed before.
\i Surprisingly, the $B$-distributions  of the isolated pulsars in the
disc and the clusters have  extremely low correlation ($P_{KS} \lsim 5
\times  10^{-4}$).   This  is  in  complete  contrast  with  the  high
correlation ($P_{KS}  \sim 0.9$)  of their $P$-distributions.   But it
should be remembered that for  the cluster pulsars only a small number
(11 out  of 61) of  objects have a  measured $B$ and hence  only those
have  been used  for this  test.   However, as  mentioned earlier  the
$P_B^{AV}$s  of  these  two   groups  are  quite  similar,  indicating
selection of pulsars with similar $P$-values.
\i Another surprise  is the very good correlation  ($P_{KS} \sim 0.7$)
between the isolated and the  binary pulsars in the clusters, again in
complete contrast  with the poor correlation ($P_{KS}  \sim 0.002$) of
their $P$-distributions.   Though once  again we need  to note  that a
very small sample of objects (11 out of 61 for the isolated and 14 out
of 68  for the  binaries) are  used for the  $B$-test compared  to the
$P$-test.   Whether this  indicates  a similar  $B$-evolution for  the
groups is difficult to say because somewhat contrarily the $P_B^{AV}$s
are different for these two groups.
\een

\begin{table}
\begin{tabular}{|l|l|r|} \hline
Accreting Sources & $P_s$ & $B$ \\ 
& ms & G \\ 
 IGR J00291+5934$^a$         & 1.67  & $< 3 \times 10^8$     \\ 
 Aql X-1 (1908+005)$^b$      & 1.82  & $\lsim 10^9$          \\ 
 XTE J1751-305$^c$           & 2.30  & $ 3-7 \times 10^8$    \\  
 SAX J1808.4-3658$^d$        & 2.49  & $1 - 5 \times 10^8$   \\  
 XTE J1814-338$^e$           & 3.18  & $8 \times 10^8$       \\ 
 XTE J0929-314$^f$           & 5.41  & $< 3 \times 10^9$     \\ 
 SWIFT J1756.9-2508$^g$      & 5.49  & $0.4 - 9 \times 10^8$ \\ 
 KS 1731-260$^h$             & 1.91  & $\lsim 7 \times 10^8$ \\  
 EXO  0748-676$^i$           & 1.81  & $\sim 1 - 2 \times 10^9$ \\ 
 $^*$MXB 1730-335$^j$        & 3.27  & $\sim 4  \times 10^8$ \\ 
&& \\ \hline
\end{tabular}
\caption[]{AMXP and burst oscillation  sources that have some estimate
of their magnetic fields.   MXB 1730-335 (oscillation frequency yet to
be modified) in NGC 6440 is the  only one in a globular cluster with a
field estimate.   The field values  are inferred upper limits  in most
cases.  The data is taken from
a) \citeN{gall05}, \ \citeN{burd07}, \ \citeN{torr08};
b) \citeN{salv03}, \ \citeN{case08};                 
c) \citeN{wijn05}, \ \citeN{papt08};                
d) \citeN{salv03}, \ \citeN{hart08}, \ \citeN{cack09};
e) \citeN{papt07};                                   
f) \citeN{gall02}, \ \citeN{wijn05};                
g) \citeN{krim07}, \ \citeN{patr10a};               
h) \citeN{salv03};  
i) \citeN{loeb03};
and j) \citeN{mast00} respectively.}
\label{tab02}
\end{table}
In summary then we can think of the following possibilities giving
rise to the higher surface fields observed in the cluster MSPs.  
\ben
\i  The  higher  field  observed   could  simply  be  due  to  certain
observational  biases   (preferentially  selecting  high-$B$  systems,
measuring a higher value of $\pdot$ and hence $B$ due to proper motion
contamination and so on..).
\i  It is  also  possible that  the  cluster MSPs  have higher  fields
because we are looking at a younger population in the cluster compared
to the  disc.  These  young MSPs observed  today would slow  down with
time  and migrate towards  the right  of the  $B-P$ plane.  Looking at
Fig.\ref{fig04} we can  see that if the cluster  MSPs move towards the
right of the plot by appropriate amount then they may become identical
with  the current  disc population.   There  is some  support to  this
possibility from the  observed AMXPs.  The AMXPs turn  into radio MSPs
as  soon as  accretion stops  in  such systems.   In principle,  their
physical properties  should be similar  to very young radio  MSPs.  In
table\ref{tab02} we  list the  AMXPs and the  burst sources  that have
some estimate  of their magnetic  fields.  These have been  plotted in
Fig.\ref{fig04} and  we see that in  the $B-P$ plane  the AMXPs occupy
the same region as the cluster MSPs.
\i On the other hand, the cluster MSPs could really have higher fields
compared  to the disc  MSPs. This  is possible  only if  the different
nature  of the  recycling process  in the  clusters affects  the field
evolution significantly.   In the next  section we use a  simple model
for the evolution of the magnetic field to see if the stellar dynamics
in globular  clusters facilitate halting the field  decay earlier than
is expected in a primordial LMXB.
\een

\section{recycling : evolution of the magnetic field}
\label{sec03}

In the standard formation scenario the MSPs are generated by recycling
of ordinary pulsars in the LMXBs.  Accretion-induced field decay is an
integral  part  of this  generic  picture.   A  number of  mechanisms,
responsible  for  the evolution  of  the  field,  has been  suggested.
First, the  magnetic field may be  dissipated in the  stellar crust by
Ohmic  decay, accelerated by  heating as  the accreted  plasma impacts
upon  the star~\cite{kb97,urpn98a,brwn98,kb99a,cumm04}.  On  the other
hand, if the magnetic field resides in the superfluid core in the form
of Abrikosov fluxoids then they may  be dragged out of the core by the
outward   motion   of  superfluid   vortices,   as   the  star   spins
down~\cite{srini90,miri94,rudr98}.    This    flux   would   then   be
subsequently      dissipated     in      the      accretion     heated
crust~\cite{kb99b,konk01b}.  Third,  the magnetic field  could also be
screened     by      accretion-induced     currents     within     the
crust~\cite{bisn74,love05}.   In particular, the  field may  be buried
under  a  mountain of  accreted  plasma  channeled  onto the  magnetic
poles~\cite{hame83,roma90,brwn98}.  When the  accreted matter is large
enough,  the  mountain   spreads  laterally,  transporting  the  polar
magnetic  flux  towards  the  equator  and  finally  dissipating  them
there~\cite{cumm01,mela01,ck02,kc04,payn04,payn07}.

In  our  earlier investigations  we  have  assumed  that the  currents
supporting the  field finally get  dissipated in the  accretion heated
crust,       wherever       they       originally       may       have
resided~\cite{kb97,kb99a,kb99b}.   We adopt the  methodology developed
in these articles (see \citeN{sk97}  for details) for the present work
and assume that purely  crustal currents support the observed magnetic
field of the neutron star.

The  evolution of  the magnetic  field  is governed  by the  following
equation
\beq
\frac{\partial \bb}{\partial t}
= \nb \times ({\bf V} \times \bb)
  - \frac{c^2}{4 \pi} \nb
    \times (\frac{1}{\sigma}
    \times \nb \times \bb),
\label{eq07}
\eeq
where {bf V} is the velocity  of material movement  and $\sigma$ is
the electrical  conductivity of the medium.  The  material movement in
the crust due to mass transfer defines {\bf V}.

The spin-up  of a neutron star, in  a binary system, is  caused by the
angular momentum brought in  by the accreted matter. In magnetospheric
accretion  matter  accretes  with  angular momentum  specific  to  the
Alfv\'{e}n radius. Therefore, the total angular momentum brought in by
accretion is :
\beq
J_{\rm accreted} = \delta M R_A V_A,
\label{eq05}
\eeq
where $\delta M$  is the total mass accreted, $R_A$  and $V_A$ are the
Alfv\'{e}n radius  and Keplerian velocity  at that radius.   The final
period of the neutron star then is :
\beq
P_{\rm final} = 2 \pi \frac{I_{\rm ns}}{J_{\rm accreted}},
\label{eq06}
\eeq
where $I_{\rm ns}$  is the moment of inertia of  the neutron star. The
average dipolar field of the disc MSPs is $\sim 10^8$~G, whereas it is
$\sim   10^9$~G    for   the   MSPs   in    globular   clusters   (see
Table-\ref{tab01}).  So the minimum mass  required to a spin a pulsar,
with a surface magnetic field of $\sim 10^9$~G, to $\sim$5~ms would be
$\delta M \sim 0.01 - 0.1$~\msun.   The radius and the crustal mass of
a neutron star remain effectively constant for this amount of accreted
masses considered, and the corresponding change in the crustal density
profile is negligible. We therefore take  the mass flux to be the same
throughout the crust, equal to its value at the surface.  Assuming the
mass flow  to be spherically symmetric  in the deep  crustal layers of
interest, one obtains the velocity of material movement to be
\beq
\vec V = - \frac {\mdot}{4\pi r^2 \rho (r)} \hat{r},
\label{eq08}
\eeq
where  \mdot\ is  the rate  of mass  accretion and  $\rho (r)$  is the
density as  a function  of radius  $r$. 

The physical conditions of  the crust enters Eq.\ref{eq07} through the
electrical conductivity,  $\sigma$, which  is a function  of -  i) the
density of the current concentration, ii) the temperature of the crust
and iii) the  impurity content of the crustal  lattice (negligible for
an  accretion heated  crust).   We  have assumed  the  currents to  be
concentrated  at  a  density  $\sim 10^{13}$~\gcc,  and  the  impurity
concentration to  be $\sim 0.01$ for  a cold crust.   These values are
consistent  with the  fact that  the magnetic  fields of  the isolated
pulsars   do   not   decay   significantly  over   a   time-scale   of
$10^6$~yrs~\cite{bhat92,hart97}.   

The temperature  of the crust  is another important parameter  for the
field  evolution.  After  the neutron  star is  born  it monotonically
cools down through copious emission of neutrinos~\cite{page98}.  When,
in the  course of binary  evolution, the neutron star  actually starts
accreting mass - the  thermal behavior changes drastically.  Accretion
releases  heat and  the crustal  temperature quickly  (in  about $\sim
10^5$~yrs) settles  down to  a more or  less uniform and  steady value
determined   by   the   accretion  rate~\cite{mira90,zdun92}.   Though
analytical expressions giving the crustal temperature for a given rate
of accretion have been obtained by \citeN{zdun92} and \citeN{urpn95c},
the  values are  too high  for temperatures  for $\mdot  \ge  2 \times
10^{-10}$~\msun/yr. Observations  of thermal black-body  radiation from
the surface of accreting neutron stars indicate that for higher \mdot\
the temperature  of the deep  crustal layers probably  saturate around
$\sim 10^8$~K  (see, for  example, \citeN{brwn09}) .   Accordingly for
$\mdot \ge 10^{-9}$~\msun/yr we  assume the crustal temperatures to be
in the range  $\sim 10^8 - 5 \times 10^8$~K.  It  should be noted that
we consider only the temperature of the deep crustal layers, where the
currents are assumed  to be concentrated.  And the  temperature of the
entire  crust beyond a  density of  $10^{10}$~\gcc is  practically the
same  though  it  drops by  almost  two  orders  of magnitude  at  the
outermost layers of the star~\cite{gudm83,pote97}. 

From the point of view  of mass transfer (responsible for the physical
process  of recycling), pulsars  go through  three distinct  phases of
evolution    in     typical    LMXBs.    These     phases    are    as
follows~\cite{bhat91,verb93,heuv95} -
\ben
\i The  Isolated Phase -  Though the stars are  gravitationally bound,
there  is no  mass transfer.   In  general, the  isolated phase  lasts
between $10^8 - 10^9$ years.
\i The Wind Phase - The interaction is through the stellar wind of the
companion which is likely to be in its main-sequence.  In general this
phase  lasts for about  $10^8 -  10^9$ years  with attendant  rates of
accretion ranging from about $10^{-15}$~\dmdt to $10^{-12}$~\dmdt.
\i The  Roche-contact Phase - When  the companion of  the neutron star
fills its Roche-lobe  a phase of heavy mass  transfer ensues.  In this
phase, the mass  transfer rate could be as high  as the Eddington rate
($10^{-8}$~\dmdt for  a 1.4 $\msun$ neutron star),  lasting for $\lsim
10^8$ years.   But there has  also been indications that  the low-mass
binaries  may even  spend $\sim  10^{10}$ years  in  the Roche-contact
phase  with a  sub-Eddington accretion  rate~\cite{hans98}.   For wide
binaries,  however, the  contact phase  may last  as little  as $10^7$
years.
\een

Earlier  we have  followed the  evolution  of the  magnetic field  for
neutron  stars  in  different  types  of binaries,  assuming  all  the
binaries  to  be  primordial,  undergoing standard  phases  of  binary
evolution.  For  our adopted model~\cite{kb97,kb99a} it  has been seen
that  the field decays  rapidly in  the initial  phase, followed  by a
slow-down  and a  final  freezing. The  initial  decay is  due to  the
heating  of  the crust  in  which  the  currents undergo  rapid  ohmic
dissipation. But as the accretion  proceeds there is addition of extra
material in  the outer layers of  the star. Since an  increase of mass
makes a neutron  star more compact, in the deeper  layers of the crust
this induces an inward radial motion. As a result the current carrying
layers progressively move into  higher density and higher conductivity
region. Consequently the decay slows down. And when the entire current
distribution,  responsible for  the field,  gets assimilated  into the
highly  conducting (time-scale  of  diffusion larger  than the  Hubble
time) core the decay stops  altogether freezing the field at its final
value.   Understandably there  is no  further evolution  of  the field
after freezing even if mass  accretion continues.  Also the higher the
accretion rate the  sooner the freezing sets in  resulting in a higher
value of the final surface field.

In globular clusters, however,  most binaries are not primordial.  For
example,  the total  observed  number of  LMXBs  in globular  clusters
exceeds  their  formation  rate  in  the disc  by  several  orders  of
magnitude,   indicating   a   dynamical   origin~\cite{clrk75}.    The
composition  of the  binary itself  may change,  even more  than once.
Dynamical interactions in the  clusters involving at least one neutron
star  are typically  of  the two  and  the three  body  types, as  the
probability of  interactions involving four  or more objects  would be
negligibly small.  Similarly, a  three body interaction is essentially
between a single star and  a binary system. Stellar interaction in the
clusters involving  a neutron  star have been  studied by a  number of
authors~\cite{krol84,rasi91,davi92,davi98,rasi00,king03}.  We  briefly
discuss below  the interactions  which play an  active role  in pulsar
recycling (see \citeNP{caml05} for a detailed review). It needs to be
mentioned that we do not consider the interactions that have no direct
effect on the mass transfer phases of a binary pulsar (for example
the 'fly-by' kind of interaction between a single star and a binary).

\subsection{Two-Body Interactions}
The interaction  could be  a close
tidal  encounter  or  a   direct  physical  collision.   However,  the
formation  of stable  binaries through  tidal encounters  is  not very
likely. Therefore,  this process is  not important for  MSP formation.
For collisional encounters there can be mainly two types of partners.
 
{\em  Main-sequence star:}  Typically such  a collision  leads  to the
complete  destruction  of  the  main-sequence star  forming  a  thick,
rapidly rotating envelope around  the pulsar.  Depending on the amount
of accretion  the pulsar may be  spun-up to millisecond  periods or it
may  be only  mildly  recycled, either  way  giving rise  to a  single
recycled pulsar.
 
{\em Red-giant  star:} Such  a collision leads  to the formation  of a
high-eccentricity   binary.   These   collisions  provide   a  natural
formation   process  for  eccentric   low-mass  binary   pulsars  with
white-dwarf companions.   If the  post-collision neutron star  - white
dwarf  binaries  retain high  eccentricities,  then  they could  decay
through gravitational-wave emission  and possibly become ultra-compact
X-ray binaries (UCXB) with $P_b  \lsim 1$~hr.  These are important for
pulsar  recycling since  a  number  of the  known  AMXPs are  actually
members  of UCXBs,  which are  also  probably the  progenitors of  the
black-widow MSPs~\cite{king03}.

\subsection{Three-Body Interactions}

When a single star interacts with a
binary (the neutron star could either be a member of the binary or the
single  object) the  result could  be one  of the  following  types of
changes to the binary :
\ben
\i  {\bf Exchange  :} In  this  kind of  encounter one  of the  binary
components is replaced  by the single star.  So  a single pulsar could
acquire  a binary  companion through  this process.   Alternatively, a
previously formed  binary pulsar could  interact with another  star or
binary.   This would lead  to a  new companion  for an  MSP, or  for a
non-recycled pulsar, or could release an MSP from a binary, creating a
single MSP.  Systems with  higher-mass companions, fast MSPs, and very
high  eccentricities, are  likely to  be the  result of  such exchange
interactions,  i.e.,  the  presently  observed  companion  was  likely
acquired  later  and  is not  the  donor  from  which the  pulsar  was
recycled.    Exchange  interactions   between  ordinary   pulsars  and
primordial  binaries also provide  a natural  way of  forming possible
progenitors of UCXBs~\cite{king03}.
\i {\bf  Disruption :} Finally,  it is possible  for the binary  to be
completely disrupted by  its interaction with the single  star. This 
would release a single pulsar. Depending on when the binary is disrupted
the pulsar could be an MSP or a mildly recycled pulsar.
\i  {\bf Multiple  Interactions  :} In  some  of the  clusters the  stellar
densities  are so  high  that the  interaction  time-scale for  either
exchange or fly-by could be  small making it possible for the binaries
to undergo  multiple interactions. Under the  circumstances, a neutron
star may actually  go through various phases of  evolution in a number
of binaries sequentially.
\een

Given the above possibilities, a number of different situations can be
envisaged  where the  phases of  accretion are  rather  different from
those  in  typical  LMXBs.  It  has  been  seen  that at  least  $\sim
10^{-2}$~\msun is required to spin a pulsar up-to millisecond periods.
We have  seen earlier that  accretion of $\sim 10^{-2}$~\msun  is also
enough for the field to  freeze to its final stable value~\cite{kb97}.
Because the  mass of the crust  of a typical 1.4\msun  neutron star is
$\sim 10^{-2}$~\msun.   And when this amount of  material is accreted,
the entire mass of the  original crust containing the current carrying
layers get accumulated into the highly conducting core where there can
be no dissipation of the field.   Once this amount of mass is accreted
the field attains its final value even though the star may continue to
go  through further accretion  and spin-up.   Below, we  list possible
situations where  $M \gsim 10^{-2}$~\msun  \ could be accreted  onto a
neutron star.  For our calculations we have evolved the field till the
final frozen value is attained.
\bef 
\epsfig{file=fig06.ps,width=165pt,angle=-90}
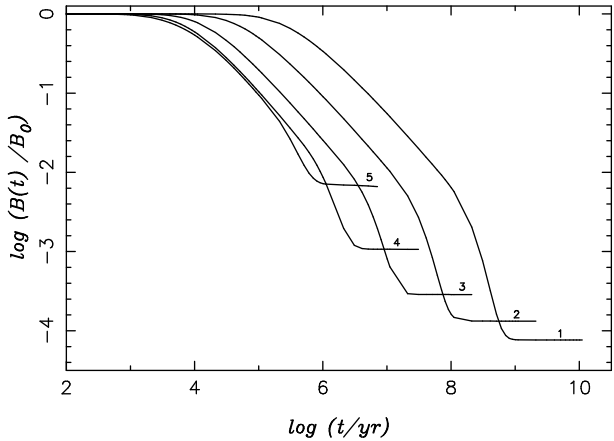
\caption[]{Evolution of  the surface magnetic field of  a neutron star
as a  result of  mass accretion, assuming  the crustal currents  to be
concentrated at a density of $10^{13}$~\gcc.  Curves 1 to 5 correspond
to  mass accretion  rates of  $\mdot =  10^{-12},  10^{-11}, 10^{-10},
10^{-9},  10^{-8}$~\msun/yr respectively.  The calculations  have been
performed till the field  decays no further.}
\label{fig06} 
\eef

\ben
\i The entire mass  is accreted with \mdot\  characteristic of
the wind phase  ($\mdot \sim 10^{-14} - 10^{-12}$)  in low-mass stars.
This situation can arise if  the original low-mass binary is disrupted
before Roche-contact is  established.  The only way a  pulsar could be
spun up-to millisecond  period entirely by wind accretion  is if $\mdot
\sim 10^{-12}  - 10^{-11}$~\msun/yr, assuming  the main-sequence phase
of an  extremely  low-mass companion  to   last  for   $\sim   10^9  -
10^{10}$~yrs. \\
\i A brief  wind phase  (or complete  absence it),  followed by
heavy   mass  transfer   ($\mdot  \sim   10^{-10}   -  10^{-8}$~\msun)
characteristic of the Roche-contact phase in which most of actual mass
accretion  takes place.   It  needs  to be  mentioned  here that  this
situation is very similar to typical LMXB evolution, except for a long
wind phase with attendant low  values of \mdot\ prior to Roche-contact
which is  realized in  most LMXBs.  In  our earlier  investigations we
have seen that the final field values attained with or without a phase
of wind accretion are not greatly different~\cite{kb99a}.  Hence, even
though the  nature of  binary evolution would  be different  the final
$P,B$ seen in the resultant MSP would be similar to one processed in a
primordial low-mass binary.  Now,  this kind of mass transfer scenario
is possible if an un-recycled  (mildly or otherwise) pulsar acquires a
partner which is already in  an evolved phase (for example a red-giant
star), in  a tight  binary.  It should  be noted that  collisions with
red-giant stars are expected to  give rise to UCXBs, extremely compact
systems that are likely to have mass transfer at high \mdot~s. \\
\i A brief phase  of heavy mass  transfer, followed by  a long
interval with  low \mdot\  , most  of the mass  being accreted  in the
second  phase.  This situation  can arise  either through  an exchange
interaction (the pulsar being always in the binary) or due to a binary
(containing a pulsar)  disruption followed by an acquisition  of a new
partner by the pulsar.  We find that this is very similar to case {\bf
A}  with  the initial  phase  of  heavy  accretion having  hardly  any
significant effect. \\
\i A  phase of heavy mass transfer followed  by a brief interval
of accretion with  small \mdot\ where most of the  mass is accreted in
the first phase.  This is again very similar to case {\bf B}.  \\
\een

In  summary, it  can  be said  that  given the  time-scales of  binary
evolution and stellar collisions the amount of mass required to freeze
the magnetic field to its  final stable value would be mostly achieved
in  one  particular  phase  of  accretion  with  a  given  \mdot.   In
Fig.\ref{fig06} we plot  the evolution of the surface  field with time
for different values of \mdot. It is seen that a higher magnetic field
is retained  by the  pulsar if most  of the  mass is transferred  in a
short  period with  a high  \mdot\,  even though  initially the  field
decays  faster  due to  a  higher  crustal  temperature. This  happens
because  the current  carrying layers  get assimilated  into  the core
(which  has effectively  infinite  conductivity) faster  for a  higher
\mdot.  A high rate of mass  transfer can be realized if the pulsar is
in a tight binary.  Interestingly, it is already known that the binary
MSPs in the clusters  have relatively shorter orbital periods (tighter
orbits) compared to the  disc population.  Dynamically too, collisions
of single  pulsars with red-giant  stars or exchange  interactions are
likely to produce UCXBs -  again systems where higher values of \mdot\
can be achieved.  So there  definitely exist MSP formation channels in
the clusters that are conducive of producing high magnetic field MSPs.
But, in  general, the  average field values  would be higher  than the
disc MSPs only  if this were the dominant channel  of MSP formation in
the clusters.  In  an ongoing work we are looking  at this question of
relative  importance of  different channels  of MSP  formation  in the
clusters (Bagchi \& Konar, in preparation) but the results are not 
available yet.

\section{conclusions}
\label{sec06}

In this  work, we  have compared  the MSPs in  the Galactic  disc with
those  in the  globular  clusters vis-vis  their  distribution of  the
spin-period and the surface  magnetic field. The statistical nature of
the two populations are as follows.
\ben
\i The  average spin-periods of  isolated and binary MSPS  in globular
clusters  as well  as  isolated MSPs  in  the Galactic  disc are  very
similar. Though the average spin-period of the binary MSPs in the disc
is somewhat different, the difference is not very significant.
\i  The  Kolmogorov-Smirnov  probabilities  indicating the  extent  of
correlation of the spin--period  between various subgroups of MSPs are
found to be as follows -
\bei
\i isolated vs. binary in the clusters - $P_{KS} \sim 10^{-3}$,
\i isolated vs. binary in the disc - $P_{KS} \sim 10^{-1}$,
\i isolated  in the clusters vs.  isolated in the disc  - $P_{KS} \sim
0.9$,
\i  binary in  the clusters  vs.  binary in  the disc  - $P_{KS}  \sim
10^{-2}$.
\eei
\i Finally, the  subset of cluster MSPs, for  which field measurements
are available, appear to have 2-5 times higher surface magnetic fields
compared to the disc MSPs. Though the MSPs with field measurements are
actuallay  a slower  subset of  the  cluster MSPs  their average  spin
periods are  similar to the  MSPs in the  disc which are,  in general,
slower than the  cluster MSPs.  We feel the  difference in the surface
magnetic field  could be  due to one  or several of  the possibilities
listed below.
\bei
\i There  are systematic biases  in \pdot\ (hence $B$)  measurement for
cluster MSPs.
\i  The cluster  MSPs are  younger and  may evolve  to  a distribution
similar to the disc MSPs with time.
\i Preferential recycling of MSPs  in tighter binaries with high rates
of  attendant mass  transfer may  actually result  in  cluster pulsars
retaining higher magnetic fields.
\eei
\een

\section{acknowledgments}

The author  wishes to thank  Manjari Bagchi and  Dipankar Bhattacharya
for  useful discussions.  The  author  is also  very  grateful to  the
anonymous  referee for  making  several valuable  comments which  have
helped improve the paper significantly.

This work has made extensive use  of the data from ATNF on-line pulsar
catalog~\cite{manc05}                      at                     {\bf
http://www.atnf.csiro.au/research/pulsar/psrcat/},  Paulo  C. Freire's
on-line    catalog    of   globular    cluster    pulsars   at    {\bf
http://www.naic.edu/~pfreire/GCpsr.html}    and    Duncan    Lorimer's
review~\cite{lorm08}  of  binary   and  millisecond  pulsars  at  {\bf
http://relativity.livingreviews.org/Articles/lrr-2008-8/}.    All  the
data taken from these websites  correspond to that available in April,
2010.

\bibliography{mnrasmnemonic,/home/sushan/prof/text/rsrc/REFS/refs}

\bibliographystyle{mnras}

\bsp

\label{lastpage}
\end{document}